\documentclass[a4paper]{article}
\usepackage{graphicx}
\usepackage[left=1.5cm, right=1.5cm, top=1.5cm, bottom=2cm]{geometry}
\usepackage{amsmath}
\usepackage{amssymb}
\usepackage{arydshln}
\usepackage{upgreek}
\usepackage{xcolor}
\usepackage{caption}
\usepackage{verbatim}
\usepackage{multicol}
\usepackage{setspace}
\usepackage{etoolbox}
\usepackage{pdfpages}

\usepackage{xr}
\usepackage{float}
\newfloat{mybox}{thp}{ext}
\floatname{mybox}{Box}

\definecolor{blue}{RGB}{0,0,255}
\definecolor{red}{RGB}{255,0,0}

\newcommand{\Ca}{Ca$^{2+}$ }
\newcommand{\Canosp}{Ca$^{2+}$}
\newcommand{\Cac}{[Ca$^{2+}$] }
\newcommand{\Cacnosp}{[Ca$^{2+}$]}
\newcommand{\K}{K$^{+}$ }
\newcommand{\Na}{Na$^{+}$ }
\newcommand{\CACNA}[2]{\textit{CACNA{#1}{#2}}}
\newcommand{\CACNB}[1]{\textit{CACNB{#1}}}
\newcommand{\ATP}[3]{\textit{ATP{#1}{#2}{#3}}}

\newcommand{\SCN}[2]{\textit{SCN{#1}{#2}}}
\newcommand{\KCNS}[1]{\textit{KCNS{#1}}}
\newcommand{\KCNN}[1]{\textit{KCNN{#1}}}
\newcommand{\HCN}[1]{\textit{HCN{#1}}}
\newcommand{\KCNB}[1]{\textit{KCNB{#1}}}

\newcommand{\KCNG}[1]{\textit{KCNG{#1}}}
\newcommand{\KCNH}[1]{\textit{KCNH{#1}}}

\title{Functional effects of schizophrenia-linked genetic variants on intrinsic single-neuron excitability: A modeling study.}
\author{Tuomo M\"aki-Marttunen$^{1}$, Geir Halnes$^{2}$, Anna Devor$^{3,4,5}$, Aree Witoelar$^{1}$, Francesco Bettella$^{1,6}$, \\
Srdjan Djurovic$^{7,8}$, Yunpeng Wang$^{1,6}$, Gaute T. Einevoll$^{2,9}$, Ole A. Andreassen$^{1,6}$, Anders M. Dale$^{3,4}$}

\date{}

\begin{document}


\maketitle
\begin{center}
$^{1}$NORMENT, KG Jebsen Centre for Psychosis Research, Institute of Clinical Medicine, University of Oslo, Oslo, Norway \\
$^{2}$Department of Mathematical Sciences and Technology, Norwegian University of Life Sciences, {\AA}s, Norway \\
$^{3}$Department of Neurosciences, University of California San Diego, La Jolla, CA, USA \\
$^{4}$Department of Radiology, University of California, San Diego, La Jolla, CA, USA \\
$^{5}$Martinos Center for Biomedical Imaging, Massachusetts General Hospital, Harvard Medical School, Charlestown, MA, USA \\
$^{6}$Division of Mental Health and Addiction, Oslo University Hospital, Oslo, Norway \\
$^{7}$Department of Medical Genetics, Oslo University Hospital, Oslo, Norway \\
$^{8}$NORMENT, KG Jebsen Centre for Psychosis Research, Department of Clinical Science, University of Bergen, Bergen, Norway \\
$^{9}$Department of Physics, University of Oslo, Oslo, Norway \\
\vspace{5pt}
Corresponding author: Anders M. Dale, 8950 VLJ Dr/C-101 La Jolla, CA, USA (amdale@ucsd.edu)\\

\end{center}



\section*{Abstract}
\textbf{Background}\\
Recent genome-wide association studies (GWAS) have identified a large number of genetic risk factors for schizophrenia (SCZ) featuring ion channels
and calcium transporters. For some of these risk factors, independent prior investigations have examined the effects of
genetic alterations on the cellular electrical excitability and calcium homeostasis.
In the present proof-of-concept study, we harnessed these experimental results
for modeling of computational properties on layer V cortical pyramidal cell and identify possible common alterations in behavior across
SCZ-related genes.\\
\textbf{Methods}\\
We applied a biophysically detailed multi-compartmental model to study the excitability of a layer V pyramidal cell.
We reviewed the literature on functional genomics for variants of genes associated with SCZ, and used changes in neuron model parameters to represent the effects of these variants.\\
\textbf{Results}\\
We present and apply a framework for examining the effects of subtle single nucleotide polymorphisms in ion channel and \Ca transporter-encoding genes on neuron excitability.
Our analysis indicates that most of the considered SCZ-related genetic variants affect the spiking behavior and intracellular calcium dynamics resulting from summation of inputs across
the dendritic tree.\\
\textbf{Conclusions}\\
Our results suggest that alteration in the ability of a single neuron to integrate the inputs and scale its excitability may
constitute a fundamental mechanistic contributor to mental disease, alongside with the previously proposed deficits in synaptic communication and network behavior.

\section*{Introduction}

Schizophrenia (SCZ) is a severe mental disorder with heritability estimates ranging from 0.6 to 0.8 \cite{ripke2013genome}.
A recent genome-wide association study (GWAS) has identified more than a hundred genes exceeding genome-wide significance, confirming the polygenic nature of
  this psychiatric disorder \cite{ripke2014biological}.
This remarkable success in gene discovery brings up the next big challenge for psychiatric genetics -- translation of the genetic associations into biological
  insights \cite{os2009schizophrenia}. 
Attaining this goal is supported by the development of biophysically detailed neuron models, boosted by the recent launch of mega-scale neuroscience projects \cite{grillner2014megascience}.
These models make it possible to investigate SCZ disease mechanisms by computational means, ultimately aiming towards achieving better clinical treatments and
  disorder outcomes \cite{wang2014computational,owen2014new}. 

The 108 recently confirmed SCZ-linked loci span a wide set of protein-coding genes \cite{ripke2014biological}, including numerous ion channel-encoding genes.
The disorder is associated with genes affecting transmembrane currents of all major ionic species, Na$^{+}$, K$^{+}$, and \Canosp.
In addition, some of the SCZ-linked genes are involved in regulation of the \Ca
concentration in the intracellular medium \cite{ripke2014biological}, which is another great contributor to excitability. 
It is thus reasonable to hypothesize that the SCZ-linked genes should have an impact on the excitability at the single neuron level.

We focus our study on cortical layer V pyramidal cells (L5PC) as a principal computational element of the cerebral circuit.
An L5PC extends throughout the cortical depth with the soma located in layer V and the apical dendrite branching into the ``apical tuft'' in layer I, and its long axon may project to non-local
  cortical and subcortical areas.
The tuft serves as an integration hub for long-distance synaptic inputs, and is often considered a biological substrate for cortical associations
  providing high-level ``context'' for low-level (e.g., sensory) inputs to the perisomatic compartment \cite{larkum2013cellular}.
Therefore, the ability of L5PC to communicate the apical inputs to the soma has been proposed as one of the mechanisms that could be impaired in the mental disease \cite{larkum2013cellular}.
In agreement with this hypothesis, recent psychiatric GWASs consistently reported association of genes coding for the subunits of voltage-gated \Ca channels as risk
  factors in SCZ and bipolar disorder \cite{sklar2011large,ripke2011genome,smoller2013identification}.

In the present proof-of-principle study, we apply a model \cite{hay2011models} of L5PC to explore how genetic variants in SCZ-linked genes affect the single-cell excitability.
We carry out our study by linking a documented effect of a genetic variant in an ion channel or \Ca transporter-encoding gene to a change in the corresponding neuron model parameter.
It should, however, be noted that information does not generally exist for the effect of single nucleotide polymorphism (SNP) variants identified through GWASs
on the biophysical parameters required for the computational models. We instead use information obtained from \textit{in vitro} studies of more extreme
genetic variations, including loss of function mutations. A central assumption of this approach is that the effects of SNP variants can be represented as scaled-down
versions of those of the more extreme variants, and that the emergence of the full psychiatric disease phenotype results from the combined effect of a large number of subtle SNP
effects \cite{gottesman1967polygenic,purcell2009common}.
A deficit in synaptic communication is likely to contribute to SCZ \cite{stefansson2009common,fromer2014novo,wen2014synaptic,ripke2014biological} but is outside the scope of the present work.

\section*{Methods and materials}

\subsection*{The L5PC model}
The multi-compartmental neuron model used in this work is based on a reconstructed morphology of a layer V thick-tufted pyramidal neuron (cell \#1 in \cite{hay2011models}).
The model includes the following ionic currents: 
Fast inactivating \Na current ($I_{Nat}$), persistent \Na current ($I_{Nap}$), non-specific cation current ($I_h$), muscarinic \K current ($I_m$),
slow inactivating \K current ($I_{Kp}$), fast inactivating \K current ($I_{Kt}$), fast non-inactivating \K current ($I_{Kv3.1}$), high-voltage-activated \Ca
current ($I_{CaHVA}$), low-voltage-activated \Ca current ($I_{CaLVA}$), small-conductance \Canosp-activated \K current ($I_{SK}$), and finally, the passive leak
current ($I_{leak})$. See Supplemental information for the model equations, and for the simulation codes, see ModelDB entry 169457 (https://senselab.med.yale.edu/ModelDB).

\subsection*{Genes included in the study}
Since we did not aim to provide a comprehensive evaluation of a representative fraction of the genetic risk factors of SCZ,
but to provide a proof of principle of the computational modeling approach, we selected the genomic loci using the following approach.
We based our study on the recent GWAS \cite{ripke2014biological}, 
which reported significantly associated SNPs that were scattered across hundreds of genes with a variety of cellular functions.
We concentrated on those genes that encoded either ionic channels or proteins contributing to transportation of intracellular \Ca ions.

We used the SNP-wise p-value data of \cite{ripke2014biological}, and for each gene of interest, determined the minimum p-value among those SNPs that were located in the considered gene.
We performed this operation for all genes encoding either subunits of voltage-gated \Canosp, K$^{+}$, or \Na channel, subunits of an SK, leak, or hyperpolarization-activated cyclic
nucleotide-gated (HCN) channel, or \Canosp-transporting ATPases. The genes \CACNA1C, \CACNB2, \CACNA1I, \ATP2A2, and \HCN1 possessed a small minimum p-value each ($p<3\times10^{-8}$) 
 --- these genes were also highlighted in the locus-oriented association analysis as performed in \cite{ripke2014biological}. In order to extend our study to explore a larger set of
genes, we used a more relaxed threshold ($p<3\times10^{-5}$) for the minimum p-value, and obtained the following genes in addition to the previously
mentioned ones: \CACNA1D, \CACNA1S, \SCN1A, \SCN7A, \SCN9A, \KCNN3, \KCNS3, \KCNB1, \KCNG2, \KCNH7, and \ATP2B2. Of these, we omitted the genes that are not relevant for the firing behavior
of an L5PC.
It should be noted that we used the SNPs reported in \cite{ripke2014biological} only to name the above SCZ-related genes, and due to lack of functional genomics data, we could not include the
actual SCZ-related SNPs in our simulation study. Instead, we searched in PubMed for functional genomic studies reporting the effects of any genetic variant of the above genes. For details,
see Supplemental information.

\section*{Results}
\subsection*{A new framework for bridging the gap between GWASs of SCZ and computational neuroscience}
\begin{figure}
\includegraphics[clip,trim=28pt 205pt 165pt 190pt,width=0.69\textwidth]{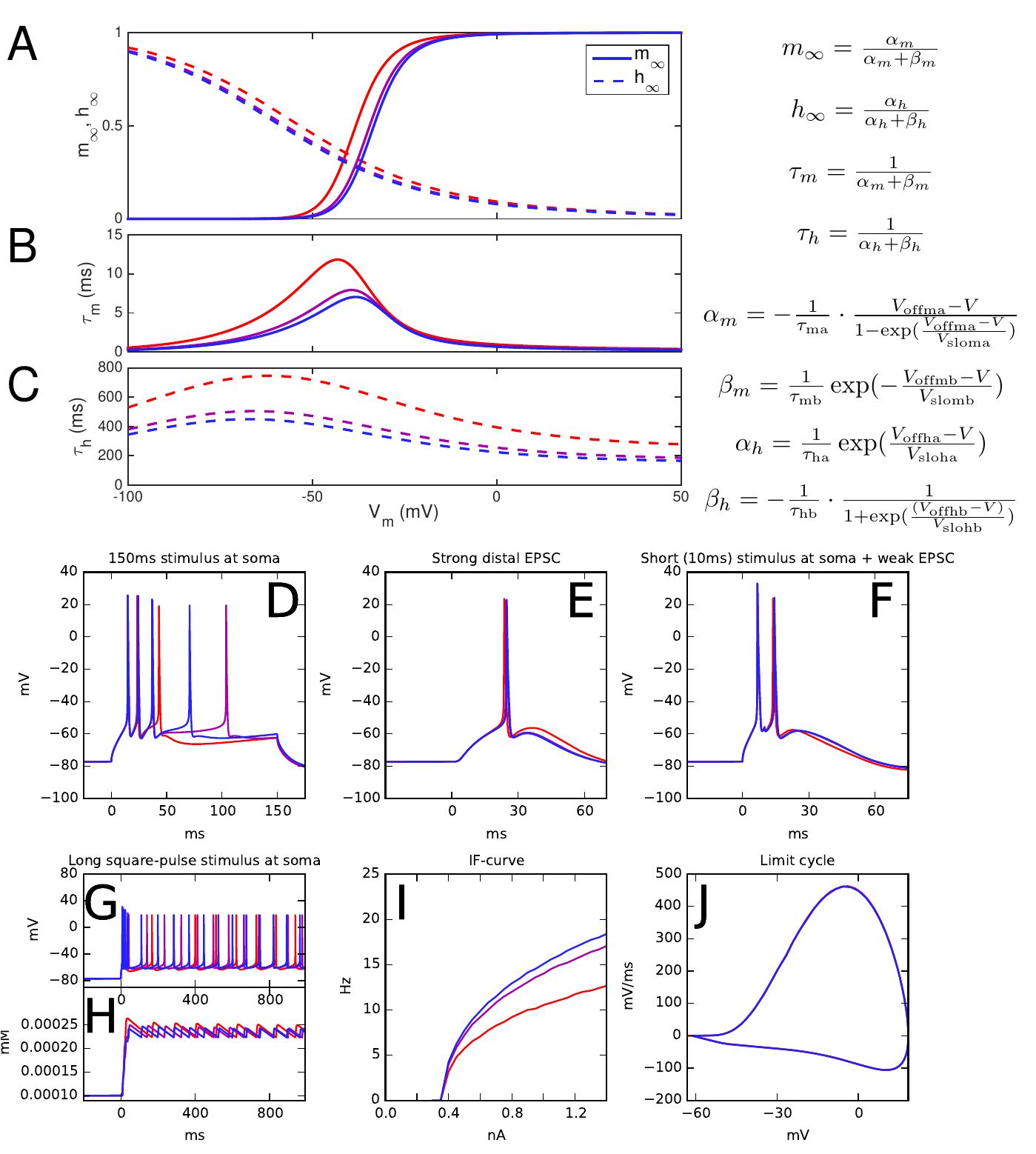}
\caption{
\textbf{An example of downscaling with a variant \cite{link2009diversity} of the gene \CACNB2}.
The authors of \cite{link2009diversity} transfected different variants of
\CACNB2 DNA into HEK293 cells together with CaV1.2 subunit DNA, and measured the activation and inactivation curves of the \Ca currents through the cell membrane.
The cells transfected with different variants showed different values of mid-points and time constants for channel activation and inactivation: The mid-points of activation
and inactivation varied by $\pm4.9$ and $\pm5.1$ mV among the variants, respectively, while the time constants of activation and inactivation varied from $-40\%$ to $+68\%$
and $-40\%$ to $+66\%$, respectively.
Here we illustrate a variant representing one possible combination of the endpoints of these ranges.
\textbf{A}: The voltage-dependence of steady-state activation ($m_{\infty}$) and inactivation ($h_{\infty}$) according to the neuron model \cite{hay2011models}.
Red curves represent the unscaled variant:
Activation parameters $V_{\mathrm{offma}}$ and $V_{\mathrm{offmb}}$ are changed by $\Delta_1 = -4.9$ mV,
inactivation parameters $V_{\mathrm{offha}}$ and $V_{\mathrm{offhb}}$ are changed by $\Delta_2 = 5.1$ mV,
activation time constants $\tau_{\mathrm{ma}}$ and $\tau_{\mathrm{mb}}$ are increased by 68\% ($\eta_3 = 1.68$),
and inactivation time constants $\tau_{\mathrm{ha}}$ and $\tau_{\mathrm{hb}}$ are increased by +66\% ($\eta_4 = 1.66$).
Purple curves show the downscaled $\epsilon=\frac 12$ variant (see below), and blue curves show the control neuron activation properties.
\textbf{B--C}: The voltage-dependence of time constants for activation (B) and inactivation (C).
\textbf{D--J}: Illustration of the scaling conditions.
The downscaling is based on five conditions (see Supplemental information), none of which should be violated in the downscaled variant.
In panels (D)--(F) (conditions I--III), the variant neuron should respond with the same number of spikes as the control neuron (blue).
In panel (I), the integrated difference between the variant and control f-I curves should not exceed 10\% of the integral of the control curve (condition IV).
In panel (J), the difference between the limit cycles should not exceed a set limit (condition V).
It can be observed that the unscaled variant (red) violates the conditions I and IV. To scale down the variant,
each applied parameter change is brought to a fraction $c < 1$, all in proportion, until the threshold of violating/non-violating variant is found. In this example, the threshold variant
corresponds to parameter changes $c\times\Delta_1$, $c\times\Delta_2$, $\eta_3^c$, and $\eta_4^c$, where $c=0.4574$. Any non-negative value below the threshold value $c$ yields a downscaled variant
that obeys conditions I--V.
The purple curves represent the variant corresponding to parameter changes
$c\epsilon\times\Delta_1$, $c\epsilon\times\Delta_2$, $\eta_3^{c\epsilon}$, and $\eta_4^{c\epsilon}$, where $c=0.4574$ and $\epsilon=\frac 12$.
The blue curves show the control neuron firing behavior.
(D): Somatic membrane potential as a response to a 150ms somatic square-pulse stimulus (0.696 nA).
(E): Somatic response to a distal apical synaptic conductance with maximum 0.0612 $\upmu$S.
(F): Somatic response to a combination of somatic square-pulse current (10 ms $\times$ 1.137 nA) and a distal apical synaptic conductance (maximum 0.100 $\upmu$S).
(G): Somatic membrane potential as a response to a long somatic square-pulse stimulus current (1.0 nA).
(H): Somatic \Ca concentration as a response to the stimulus used in (G).
(I): Spiking frequency as a function of amplitude of the somatic DC.
(J): The membrane potential limit cycle corresponding to the late phase of (G). The x-axis represents the somatic membrane potential and y-axis its time derivative.
}
\label{box1}
\end{figure}

In this work, we reviewed the literature on effects of variants in SCZ-related genes on ion channel behavior and intracellular \Ca dynamics, and interpreted the reported effects
in the context of our neuron model parameters.
An overview of the relevant studies is given in Table \ref{tab_mutations}, while the effects of each variant on the L5PC model parameters are given in Table \ref{tab_params}.
These data gave us a direct interface for linking a change in the genomic data, such as a SNP or an alternative splicing, into a change in neuron dynamics.
The reported data, however, often corresponded to variants with large phenotypic consequences that in general are absent in SCZ patients. To simulate subtle cellular effects caused
by the common SNPs related to SCZ \cite{lee2012estimating}, we downscaled the variants of Table \ref{tab_params} by bringing the changed parameters closer to the control neuron
values when the reported change caused too large an effect in the neuron firing behavior. Our approach is illustrated in Figure \ref{box1}.

\begin{table}[!ht]
\caption{
\textbf{Table of the genetic variants used in this study.} For more details, see Table \ref{tab_params}.}
\begin{small}
\begin{tabular}{llll}
Gene    & Refs. &   Type of variant                           &  Cell type      \\ 
\hline
\CACNA1C & \cite{kudrnac2009coupled}                             &  L429T, L434T, S435T, S435A, S435P        &  TSA201            \\
\CACNA1C & \cite{kudrnac2009coupled}                             &  L779T, I781T, I781P                      &  TSA201           \\ 
\CACNB2  & \cite{cordeiro2009accelerated}                        &  T11I-mutant                              &  TSA201           \\ 
\CACNB2  & \cite{massa1995comparison}                            &  A1B2 vs A1 alone                         &  HEK293           \\ 
\CACNB2  & \cite{link2009diversity}                              &  N1 vs N3 vs N4 vs N5                     &  HEK293           \\ 
\CACNA1D & \cite{tan2011functional}, \cite{bock2011functional}   &  42A splices transfected             &  TSA201/HEK293    \\ 
\CACNA1D & \cite{tan2011functional}, \cite{bock2011functional}   &  43S splices transfected             &  TSA201/HEK293    \\ 
\CACNA1D & \cite{zhang2011expression},                           &  Homozyg. knockout                        &  AV node cells /  \\ 
         & \cite{perez-alvarez2011different}                     &                                           &  chromaffin cells \\ 
\CACNA1I & \cite{murbartian2004functional}                       &  Alternative splicing of exons 9 and 33   &  HEK293           \\ 
\ATP2A2  & \cite{ji2000disruption}                               &  Heter. null mutation                     &  myocytes         \\ 
\ATP2B2  & \cite{fakira2012purkinje}                             &  Heter. knockout                          &  Purkinje cells   \\ 
\ATP2B2  & \cite{empson2010role}                                 &  Homozyg. knockout                        &  Purkinje cells   \\ 
\ATP2B2  & \cite{ficarella2007functional}                        &  G283S-,G293S-mutant                      &  CHO cells        \\ 
\SCN1A   & \cite{cestele2008self}                                &  FHM mutation Q1489K                      &  Cultured neocortical cell      \\
\SCN1A   & \cite{vanmolkot2007novel}                             &  FHM mutation L1649Q                      &  TSA201         \\ 
\SCN9A   & \cite{estacion2011intra}                              &  I228M NaV1.7 variant                     &  HEK293         \\ 
\SCN9A   & \cite{estacion2008nav1.7}                             &  A1632E NaV1.7 mutation                   &  HEK293         \\ 
\SCN9A   & \cite{han2006sporadic}                                &  L858F NaV1.7 mutation                    &  HEK293         \\ 
\SCN9A   & \cite{dib-hajj2005gain}                               &  F1449V NaV1.7 mutation                   &  HEK293         \\ 
\KCNS3   & \cite{shepard1999electrically}                        &  hKv2.1-(G4S)3-hKv9 fusion inserted       &  HEK293         \\ 
\KCNB1   & \cite{bocksteins2011functional}                       &  T203K, T203D, S347K, S347D, T203W, S347W &  LTK-           \\
\KCNN3   & \cite{wittekindt2004apamin}                           &  hSK3\_ex4 isoform                        &  TSA            \\ 
\HCN1    & \cite{ishii2007tryptophan}                            &  D135W, D135H, D135N mutants              &  HEK293         \\ 
\label{tab_mutations}
\end{tabular}
\end{small}
\end{table}

\begin{table}[!h]
\caption{\textbf{List of variants and their threshold effect coefficients $c$}. The variants are ordered as in Table \ref{tab_mutations}, but the variants where
several combinations of parameter range end points were considered are listed as separate variants (see Supplemental information). The variants marked with $\dagger$
were used in Figures \ref{fig1}--\ref{fig5}, the variants marked with $\S$ were used in Figure S3, and the variant marked with $\ddagger$ was used in Figure \ref{box1}.}
\label{tab_params}
\begin{scriptsize}
\begin{tabular}{lll}
Gene & Effect on model parameters & Thresh. effect \\ \hline
\hline \CACNA1C & $V_{\mathrm{offm\ast,CaHVA}}$: -25.9 mV; $V_{\mathrm{offh\ast,CaHVA}}$: -27.0 mV & $c = $ 0.094\\
\hline \CACNA1C & $V_{\mathrm{offm\ast,CaHVA}}$: -37.3 mV; $V_{\mathrm{offh\ast,CaHVA}}$: -30.0 mV & $c = $ 0.060 $\dagger$\\
\hline \CACNB2 & $V_{\mathrm{offh\ast,CaHVA}}$: -5.2 mV; $V_{\mathrm{sloh\ast,CaHVA}}$: $\times$0.69 & $c = $ 0.582\\
\hline \CACNB2 & $\tau_{\mathrm{h\ast,CaHVA}}$: $\times$1.7 & $c \geq $ 2.000\\
\hline \CACNB2 & $V_{\mathrm{offm\ast,CaHVA}}$: -4.9 mV; $V_{\mathrm{offh\ast,CaHVA}}$: -5.1 mV; $\tau_{\mathrm{m\ast,CaHVA}}$: $\times$0.6; $\tau_{\mathrm{h\ast,CaHVA}}$: $\times$0.6 & $c = $ 0.310\\
 & $V_{\mathrm{offm\ast,CaHVA}}$: +4.9 mV; $V_{\mathrm{offh\ast,CaHVA}}$: -5.1 mV; $\tau_{\mathrm{m\ast,CaHVA}}$: $\times$0.6; $\tau_{\mathrm{h\ast,CaHVA}}$: $\times$0.6 & $c = $ 0.154\\
 & $V_{\mathrm{offm\ast,CaHVA}}$: -4.9 mV; $V_{\mathrm{offh\ast,CaHVA}}$: +5.1 mV; $\tau_{\mathrm{m\ast,CaHVA}}$: $\times$0.6; $\tau_{\mathrm{h\ast,CaHVA}}$: $\times$0.6 & $c = $ 0.205\\
 & $V_{\mathrm{offm\ast,CaHVA}}$: +4.9 mV; $V_{\mathrm{offh\ast,CaHVA}}$: +5.1 mV; $\tau_{\mathrm{m\ast,CaHVA}}$: $\times$0.6; $\tau_{\mathrm{h\ast,CaHVA}}$: $\times$0.6 & $c = $ 0.675\hspace{9pt}\S\\
 & $V_{\mathrm{offm\ast,CaHVA}}$: -4.9 mV; $V_{\mathrm{offh\ast,CaHVA}}$: -5.1 mV; $\tau_{\mathrm{m\ast,CaHVA}}$: $\times$1.68; $\tau_{\mathrm{h\ast,CaHVA}}$: $\times$0.6 & $c = $ 0.814\\
 & $V_{\mathrm{offm\ast,CaHVA}}$: +4.9 mV; $V_{\mathrm{offh\ast,CaHVA}}$: -5.1 mV; $\tau_{\mathrm{m\ast,CaHVA}}$: $\times$1.68; $\tau_{\mathrm{h\ast,CaHVA}}$: $\times$0.6 & $c = $ 0.057\\
 & $V_{\mathrm{offm\ast,CaHVA}}$: -4.9 mV; $V_{\mathrm{offh\ast,CaHVA}}$: +5.1 mV; $\tau_{\mathrm{m\ast,CaHVA}}$: $\times$1.68; $\tau_{\mathrm{h\ast,CaHVA}}$: $\times$0.6 & $c = $ 0.517\\
 & $V_{\mathrm{offm\ast,CaHVA}}$: +4.9 mV; $V_{\mathrm{offh\ast,CaHVA}}$: +5.1 mV; $\tau_{\mathrm{m\ast,CaHVA}}$: $\times$1.68; $\tau_{\mathrm{h\ast,CaHVA}}$: $\times$0.6 & $c = $ 0.078\\
 & $V_{\mathrm{offm\ast,CaHVA}}$: -4.9 mV; $V_{\mathrm{offh\ast,CaHVA}}$: -5.1 mV; $\tau_{\mathrm{m\ast,CaHVA}}$: $\times$0.6; $\tau_{\mathrm{h\ast,CaHVA}}$: $\times$1.66 & $c = $ 0.275\\
 & $V_{\mathrm{offm\ast,CaHVA}}$: +4.9 mV; $V_{\mathrm{offh\ast,CaHVA}}$: -5.1 mV; $\tau_{\mathrm{m\ast,CaHVA}}$: $\times$0.6; $\tau_{\mathrm{h\ast,CaHVA}}$: $\times$1.66 & $c = $ 0.188\\
 & $V_{\mathrm{offm\ast,CaHVA}}$: -4.9 mV; $V_{\mathrm{offh\ast,CaHVA}}$: +5.1 mV; $\tau_{\mathrm{m\ast,CaHVA}}$: $\times$0.6; $\tau_{\mathrm{h\ast,CaHVA}}$: $\times$1.66 & $c = $ 0.190\\
 & $V_{\mathrm{offm\ast,CaHVA}}$: +4.9 mV; $V_{\mathrm{offh\ast,CaHVA}}$: +5.1 mV; $\tau_{\mathrm{m\ast,CaHVA}}$: $\times$0.6; $\tau_{\mathrm{h\ast,CaHVA}}$: $\times$1.66 & $c = $ 1.687\\
 & $V_{\mathrm{offm\ast,CaHVA}}$: -4.9 mV; $V_{\mathrm{offh\ast,CaHVA}}$: -5.1 mV; $\tau_{\mathrm{m\ast,CaHVA}}$: $\times$1.68; $\tau_{\mathrm{h\ast,CaHVA}}$: $\times$1.66 & $c = $ 0.707\\
 & $V_{\mathrm{offm\ast,CaHVA}}$: +4.9 mV; $V_{\mathrm{offh\ast,CaHVA}}$: -5.1 mV; $\tau_{\mathrm{m\ast,CaHVA}}$: $\times$1.68; $\tau_{\mathrm{h\ast,CaHVA}}$: $\times$1.66 & $c = $ 0.061\\
 & $V_{\mathrm{offm\ast,CaHVA}}$: -4.9 mV; $V_{\mathrm{offh\ast,CaHVA}}$: +5.1 mV; $\tau_{\mathrm{m\ast,CaHVA}}$: $\times$1.68; $\tau_{\mathrm{h\ast,CaHVA}}$: $\times$1.66 & $c = $ 0.457 $\dagger$ $\ddagger$\\
 & $V_{\mathrm{offm\ast,CaHVA}}$: +4.9 mV; $V_{\mathrm{offh\ast,CaHVA}}$: +5.1 mV; $\tau_{\mathrm{m\ast,CaHVA}}$: $\times$1.68; $\tau_{\mathrm{h\ast,CaHVA}}$: $\times$1.66 & $c = $ 0.084\\
\hline \CACNA1D & $V_{\mathrm{offm\ast,CaHVA}}$: -10.9 mV; $V_{\mathrm{slom\ast,CaHVA}}$: $\times$0.73; $V_{\mathrm{offh\ast,CaHVA}}$: -3.0 mV; $V_{\mathrm{sloh\ast,CaHVA}}$: $\times$0.81; $\tau_{\mathrm{h\ast,CaHVA}}$: $\times$1.25 & $c = $ 0.118\\
 & $V_{\mathrm{offm\ast,CaHVA}}$: -10.9 mV; $V_{\mathrm{slom\ast,CaHVA}}$: $\times$0.73; $V_{\mathrm{offh\ast,CaHVA}}$: +3.5 mV; $V_{\mathrm{sloh\ast,CaHVA}}$: $\times$0.81; $\tau_{\mathrm{h\ast,CaHVA}}$: $\times$1.25 & $c = $ 0.106\\
\hline \CACNA1D & $V_{\mathrm{offm\ast,CaHVA}}$: -10.6 mV; $V_{\mathrm{slom\ast,CaHVA}}$: $\times$0.8; $V_{\mathrm{offh\ast,CaHVA}}$: -5.3 mV; $V_{\mathrm{sloh\ast,CaHVA}}$: $\times$0.66; $\tau_{\mathrm{h\ast,CaHVA}}$: $\times$0.72 & $c = $ 0.114\\
 & $V_{\mathrm{offm\ast,CaHVA}}$: +3.4 mV; $V_{\mathrm{slom\ast,CaHVA}}$: $\times$0.8; $V_{\mathrm{offh\ast,CaHVA}}$: -5.3 mV; $V_{\mathrm{sloh\ast,CaHVA}}$: $\times$0.66; $\tau_{\mathrm{h\ast,CaHVA}}$: $\times$0.72 & $c = $ 1.962\\
 & $V_{\mathrm{offm\ast,CaHVA}}$: -10.6 mV; $V_{\mathrm{slom\ast,CaHVA}}$: $\times$1.12; $V_{\mathrm{offh\ast,CaHVA}}$: -5.3 mV; $V_{\mathrm{sloh\ast,CaHVA}}$: $\times$0.66; $\tau_{\mathrm{h\ast,CaHVA}}$: $\times$0.72 & $c = $ 0.131\\
 & $V_{\mathrm{offm\ast,CaHVA}}$: +3.4 mV; $V_{\mathrm{slom\ast,CaHVA}}$: $\times$1.12; $V_{\mathrm{offh\ast,CaHVA}}$: -5.3 mV; $V_{\mathrm{sloh\ast,CaHVA}}$: $\times$0.66; $\tau_{\mathrm{h\ast,CaHVA}}$: $\times$0.72 & $c = $ 0.601\\
 & $V_{\mathrm{offm\ast,CaHVA}}$: -10.6 mV; $V_{\mathrm{slom\ast,CaHVA}}$: $\times$0.8; $V_{\mathrm{offh\ast,CaHVA}}$: +1.2 mV; $V_{\mathrm{sloh\ast,CaHVA}}$: $\times$0.66; $\tau_{\mathrm{h\ast,CaHVA}}$: $\times$0.72 & $c = $ 0.100\\
 & $V_{\mathrm{offm\ast,CaHVA}}$: +3.4 mV; $V_{\mathrm{slom\ast,CaHVA}}$: $\times$0.8; $V_{\mathrm{offh\ast,CaHVA}}$: +1.2 mV; $V_{\mathrm{sloh\ast,CaHVA}}$: $\times$0.66; $\tau_{\mathrm{h\ast,CaHVA}}$: $\times$0.72 & $c = $ 0.645\\
 & $V_{\mathrm{offm\ast,CaHVA}}$: -10.6 mV; $V_{\mathrm{slom\ast,CaHVA}}$: $\times$1.12; $V_{\mathrm{offh\ast,CaHVA}}$: +1.2 mV; $V_{\mathrm{sloh\ast,CaHVA}}$: $\times$0.66; $\tau_{\mathrm{h\ast,CaHVA}}$: $\times$0.72 & $c = $ 0.116\\
 & $V_{\mathrm{offm\ast,CaHVA}}$: +3.4 mV; $V_{\mathrm{slom\ast,CaHVA}}$: $\times$1.12; $V_{\mathrm{offh\ast,CaHVA}}$: +1.2 mV; $V_{\mathrm{sloh\ast,CaHVA}}$: $\times$0.66; $\tau_{\mathrm{h\ast,CaHVA}}$: $\times$0.72 & $c = $ 1.117\\
\hline \CACNA1D & $V_{\mathrm{offm\ast,CaHVA}}$: +6.6 mV; $V_{\mathrm{slom\ast,CaHVA}}$: $\times$0.75; $\tau_{\mathrm{h\ast,CaHVA}}$: $\times$0.5 & $c = $ 0.104\\
 & $V_{\mathrm{offm\ast,CaHVA}}$: +6.6 mV; $V_{\mathrm{slom\ast,CaHVA}}$: $\times$1.19; $\tau_{\mathrm{h\ast,CaHVA}}$: $\times$0.5 & $c = $ 0.068\\
 & $V_{\mathrm{offm\ast,CaHVA}}$: +6.6 mV; $V_{\mathrm{slom\ast,CaHVA}}$: $\times$0.75; $\tau_{\mathrm{h\ast,CaHVA}}$: $\times$1.12 & $c = $ 0.115\hspace{9pt}\S\\
 & $V_{\mathrm{offm\ast,CaHVA}}$: +6.6 mV; $V_{\mathrm{slom\ast,CaHVA}}$: $\times$1.19; $\tau_{\mathrm{h\ast,CaHVA}}$: $\times$1.12 & $c = $ 0.072\\
\hline \CACNA1I & $V_{\mathrm{offma,CaLVA}}$: +1.3 mV; $V_{\mathrm{offha,CaLVA}}$: +1.6 mV; $\tau_{\mathrm{m\ast,CaLVA}}$: $\times$0.87; $\tau_{\mathrm{h\ast,CaLVA}}$: $\times$0.8 & $c \geq 2.000$ \\
 & $V_{\mathrm{offma,CaLVA}}$: +1.3 mV; $V_{\mathrm{offha,CaLVA}}$: +1.6 mV; $\tau_{\mathrm{m\ast,CaLVA}}$: $\times$1.45; $\tau_{\mathrm{h\ast,CaLVA}}$: $\times$0.8 & $c \geq 2.000$ $\dagger$\hspace{3pt}\S\\
\hline \ATP2A2 & $\gamma_{\mathrm{CaDynamics}}$: $\times$0.6 & $c = $ 0.093\hspace{9pt}\S\\
\hline \ATP2B2 & $\tau_{\mathrm{decay,CaDynamics}}$: $\times$1.97 & $c = $ 0.218\\
\hline \ATP2B2 & $\tau_{\mathrm{decay,CaDynamics}}$: $\times$1.5; $c_{\mathrm{min,CaDynamics}}$: $\times$1.4 & $c = $ 0.215 $\dagger$\\
\hline \ATP2B2 & $\tau_{\mathrm{decay,CaDynamics}}$: $\times$4.45 & $c = $ 0.099\\
\hline \SCN1A & $V_{\mathrm{offm,Nat}}$: -0.3 mV; $V_{\mathrm{offh,Nat}}$: +5 mV; $V_{\mathrm{slom,Nat}}$: $\times$1.15; $V_{\mathrm{sloh,Nat}}$: $\times$1.23 & $c = $ 0.056 $\dagger$\\
\hline \SCN1A & $V_{\mathrm{offm,Nat}}$: +2.8 mV; $V_{\mathrm{offh,Nat}}$: +9.6 mV; $V_{\mathrm{slom,Nat}}$: $\times$0.984; $V_{\mathrm{sloh,Nat}}$: $\times$1.042 & $c = $ 0.069\\
\hline \SCN9A & $V_{\mathrm{offh\ast,Nap}}$: +6.8 mV & $c \geq 2.000$ \\
\hline \SCN9A & $V_{\mathrm{offh\ast,Nap}}$: +3.5 mV; $V_{\mathrm{sloh,Nap}}$: $\times$0.55; $V_{\mathrm{offm,Nat}}$: -7.1 mV; $V_{\mathrm{offh,Nat}}$: +17.0 mV; $V_{\mathrm{sloh,Nat}}$: $\times$0.69 & $c = $ 0.026\\
\hline \SCN9A & $V_{\mathrm{offm,Nat}}$: -9.1 mV; $V_{\mathrm{offh,Nat}}$: +3.1 mV & $c = $ 0.043\\
\hline \SCN9A & $V_{\mathrm{offm,Nat}}$: -7.6 mV; $V_{\mathrm{offh,Nat}}$: +4.3 mV & $c = $ 0.043\\
\hline \KCNS3 & $\tau_{\mathrm{m\ast,Kp}}$: $\times$2.0; $\tau_{\mathrm{h\ast,Kp}}$: $\times$2.5; $V_{\mathrm{sloh,Kp}}$: $\times$0.5 & $c \geq 2.000$ \\
\hline \KCNB1 & $V_{\mathrm{offm,Kp}}$: +5 mV; $V_{\mathrm{offh,Kp}}$: +3 mV; $V_{\mathrm{slom,Kp}}$: $\times$1.11; $V_{\mathrm{sloh,Kp}}$: $\times$0.86; $\tau_{\mathrm{m\ast,Kp}}$: $\times$0.5; $\tau_{\mathrm{h\ast,Kp}}$: $\times$0.53 & $c \geq 2.000$ \\
\hline \KCNB1 & $V_{\mathrm{offm,Kp}}$: +1 mV; $V_{\mathrm{offh,Kp}}$: -6 mV; $V_{\mathrm{slom,Kp}}$: $\times$1.22; $V_{\mathrm{sloh,Kp}}$: $\times$1.0; $\tau_{\mathrm{m\ast,Kp}}$: $\times$0.89; $\tau_{\mathrm{h\ast,Kp}}$: $\times$1.13 & $c \geq 2.000$ \\
\hline \KCNB1 & $V_{\mathrm{offm,Kp}}$: +6 mV; $V_{\mathrm{offh,Kp}}$: -8 mV; $V_{\mathrm{slom,Kp}}$: $\times$1.33; $V_{\mathrm{sloh,Kp}}$: $\times$1.0; $\tau_{\mathrm{m\ast,Kp}}$: $\times$0.5; $\tau_{\mathrm{h\ast,Kp}}$: $\times$0.87 & $c \geq 2.000$ \\
\hline \KCNB1 & $V_{\mathrm{offm,Kp}}$: -28 mV; $V_{\mathrm{offh,Kp}}$: -27 mV; $V_{\mathrm{slom,Kp}}$: $\times$1.11; $V_{\mathrm{sloh,Kp}}$: $\times$0.71; $\tau_{\mathrm{m\ast,Kp}}$: $\times$1.13; $\tau_{\mathrm{h\ast,Kp}}$: $\times$2.27 & $c \geq 2.000$ \\
\hline \KCNB1 & $V_{\mathrm{offm,Kp}}$: +14 mV; $V_{\mathrm{offh,Kp}}$: -21 mV; $V_{\mathrm{slom,Kp}}$: $\times$2.0; $V_{\mathrm{sloh,Kp}}$: $\times$1.0; $\tau_{\mathrm{m\ast,Kp}}$: $\times$0.39; $\tau_{\mathrm{h\ast,Kp}}$: $\times$1.2 & $c \geq 2.000$ \\
\hline \KCNB1 & $V_{\mathrm{offm,Kp}}$: -13 mV; $V_{\mathrm{offh,Kp}}$: -13 mV; $V_{\mathrm{slom,Kp}}$: $\times$1.33; $V_{\mathrm{sloh,Kp}}$: $\times$0.71; $\tau_{\mathrm{m\ast,Kp}}$: $\times$0.95; $\tau_{\mathrm{h\ast,Kp}}$: $\times$5.13 & $c \geq 2.000$ \\
\hline \KCNN3 & $c_{\mathrm{off,SK}}$: $\times$0.86; $c_{\mathrm{slo,SK}}$: $\times$1.24 & $c = $ 1.715 $\dagger$\hspace{3pt}\S\\
\hline \HCN1 & $V_{\mathrm{offma,Ih}}$, $V_{\mathrm{offmb,Ih}}$: -26.5 mV; $V_{\mathrm{sloma,Ih}}$, $V_{\mathrm{slomb,Ih}}$: $\times$0.64 & $c = $ 0.296\\
\end{tabular}
\end{scriptsize}
\end{table}

The variants in Table \ref{tab_mutations} were 23 in total, although some of them represented a range of effects of different variants (see Supplemental information, Table S1). The entries
corresponded to variants of genes encoding for \Ca channel subunits (\CACNA1C, \CACNB2, \CACNA1D, \CACNA1I), intracellular \Ca pumps (\ATP2A2, \ATP2B2),
\Na channel subunits (\SCN1A, \SCN9A), \K channel subunits (\KCNS3, \KCNB1, \KCNN3), and a non-specific ion channel subunit (\HCN1). In the following, we
present simulation results for the L5PC model equipped with some of these downscaled variants (we refer to these model neurons as ``variant neurons'' or simply as
``variants''). As we do not know the quality of the effects of the actual SCZ-related polymorphisms, we perform the simulations for a range of differently scaled
variants, including negative scalings (i.e. opposite effects w.r.t. the effects reported in Table \ref{tab_params}).
We concentrated our study on a representative sample of six variants, highlighted in Table \ref{tab_params}. These variants represent six genes with different roles
in L5PC electrogenesis and a wide range of observed effects (see Table S2).

\subsection*{Variants show altered intracellular \Cac responses to short stimuli}
\begin{figure}[!ht]
\centering
\includegraphics[clip,trim=45pt 480pt 55pt 25pt,width=0.8\textwidth]{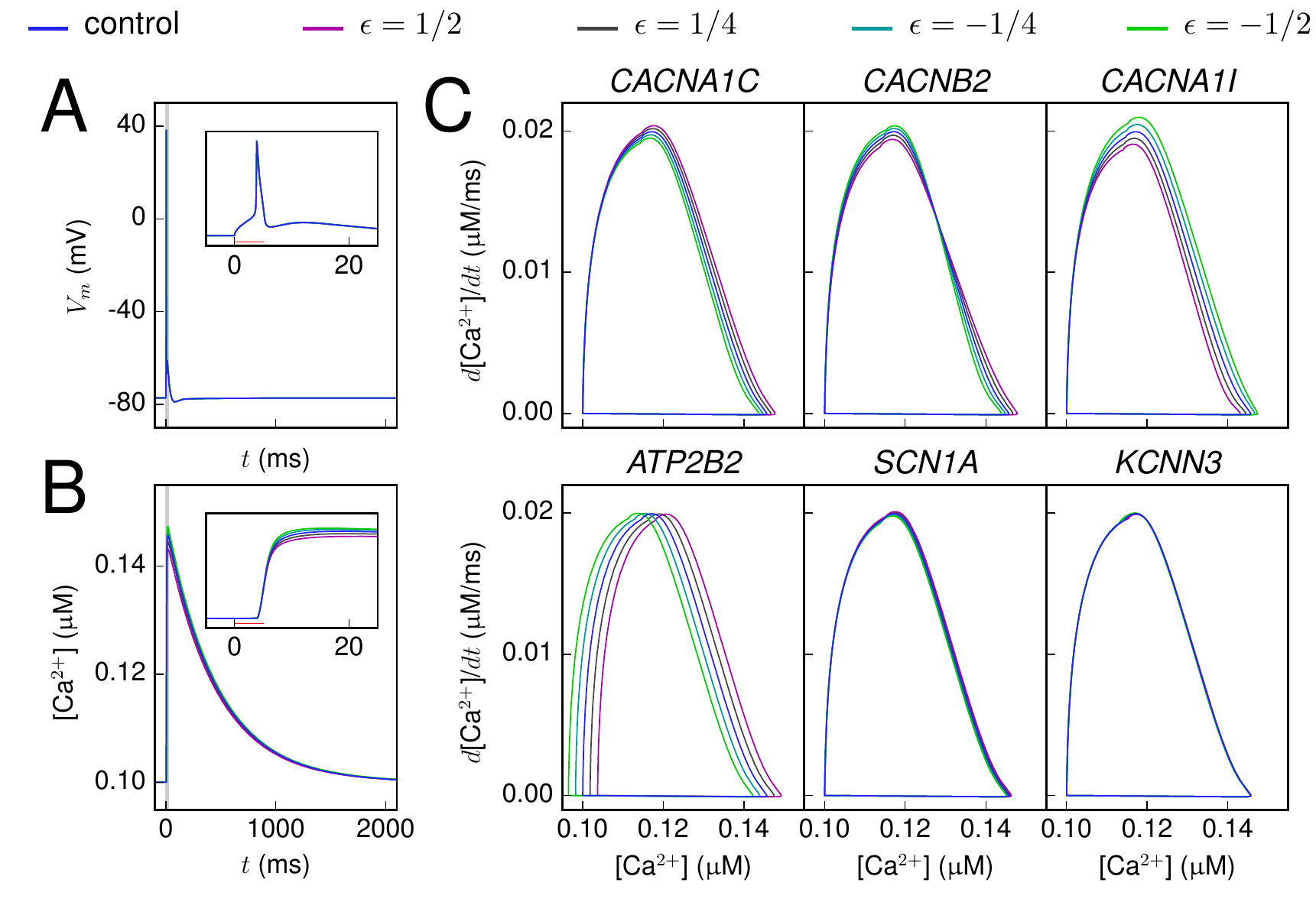}
\caption{
\textbf{Variants show altered \Cac response to a short, somatic stimulus.} \textbf{A,B}: The membrane potential (A) and \Ca concentration (B) time courses, recorded at soma, as a response to
a somatic 5 ms, 1.626 nA square-pulse stimulus. The blue curve shows the control neuron behavior, while the other curves show the behavior of a 
\CACNA1I variant \cite{murbartian2004functional} with different scalings
(magenta: $\epsilon=\frac 12$, dark gray: $\epsilon=\frac 14$, cyan: $\epsilon=-\frac 14$, green: $\epsilon=-\frac 12$). Inset: Zoomed-in view around the time of the spike. No notable differences between
the variants and the control can be observed in the membrane potential time course. \textbf{C}: The \Cac response plotted in the phase plane for different variants. The behavior of variants of \CACNA1C
\cite{kudrnac2009coupled}, \CACNB2 \cite{link2009diversity}, \CACNA1I \cite{murbartian2004functional}, \ATP2B2 \cite{empson2010role}, \SCN1A \cite{cestele2008self}, and \KCNN3 \cite{wittekindt2004apamin}
genes are shown with similar scaling as in (A) and (B).
}
\label{fig1}
\end{figure}

To characterize the implications of SCZ-related genes on the neuron excitability, we started by analyzing the effects of the downscaled versions of genetic variants
in Table \ref{tab_params} on the neuron response to a short somatic suprathreshold square-pulse stimulus. Figure \ref{fig1}A-B shows the time course of the membrane potential
and the intracellular \Canosp-concentration (\Cacnosp) for one variant, whereas Figure \ref{fig1}C shows the \Cac response in the phase plane for several different variants.
The most typical effect of a variant was a deviation in the peak \Cacnosp, but differences could also be observed in the rising phase of the \Cacnosp, as shown in the
phase plane representation in Figure \ref{fig1}C. 

The results in Figure \ref{fig1} show that the \Cac response served as a more sensitive indicator of genetic effects than the membrane potential. In \CACNA1C and \CACNB2
variants, which affect the high-voltage-activated (HVA) \Ca current, and in \CACNA1I variant affecting the low-voltage-activated (LVA) \Ca current, the observed effects on the \Cac response
were due to changed \Ca channel kinetics. Similarly, in the \ATP2B2 variant, which affects the plasma membrane \Ca ATPase (PMCA) activity, the observed effect was caused
by altered intracellular \Ca dynamics. By contrast, the small yet observable differences between control and SCN1A variant \Cac phase planes were due to alterations
in subthreshold membrane potential fluctuations, which caused variation in the activation of \Ca channels. 

\subsection*{Steady-state firing is influenced by the variants}
\begin{figure}[!ht]
\centering
\includegraphics[clip,trim=50pt 470pt 55pt 25pt,width=0.6\textwidth]{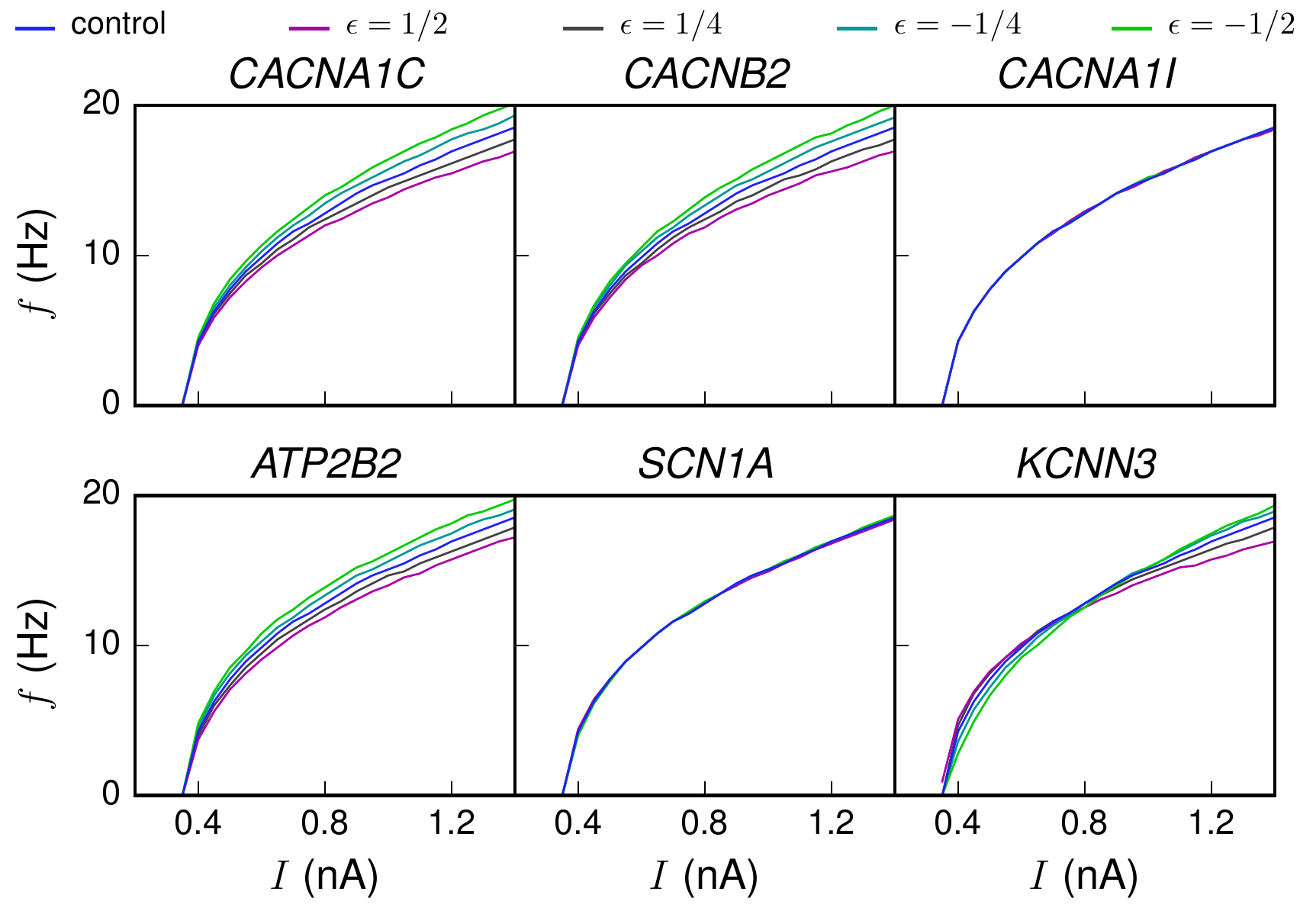}
\caption{
\textbf{The variant neurons show modulated gain.} f-I curves are shown for different scalings of different variants
(magenta: $\epsilon=\frac 12$, dark gray: $\epsilon=\frac 14$, cyan: $\epsilon=-\frac 14$, green: $\epsilon=-\frac 12$). Differences
in gain are visible for \CACNA1C, \CACNB2, \ATP2B2, and \KCNN3 variants.
}
\label{fig2}
\end{figure}

Next, we investigated the steady-state behavior of the neuron when a direct current (DC) was applied to the soma. As shown in Figure \ref{fig2}, the f-I curves (firing frequency
as a function of DC amplitude) of many SCZ-associated variants were notably different from those obtained with the control neuron.

The deviations in the f-I curves in many of the variants can be explained by changes in the \Canosp-activated SK current. These changes were caused either by
direct alteration of the activation kinetics of SK channels (as in the \KCNN3 variant), or indirectly, through the changes in the intracellular \Cac response
associated with other variants (\CACNA1C, \CACNB2 and \ATP2B2). As an example, the \Ca influx
during an action potential was larger in the \CACNA1C variant neuron compared to the control neuron (compare curves in blue and magenta in Figure \ref{fig1}C), and this
led to an increase in the SK current activation, relative to that in the control neuron. The increased $I_{SK}$, in turn, delayed the induction of the next spike, and hence
resulted in a loss of gain (flattening) in the f-I curve (Figure \ref{fig2}). This finding is in line with previous modeling studies, which also have highlighted the
role of \Canosp-activated \K currents in modulating f-I curves \cite{engel1999small, halnes2011multi}.

\begin{figure}[!ht]
\centering
\includegraphics[clip,trim=50pt 479pt 55pt 45pt,width=0.8\textwidth]{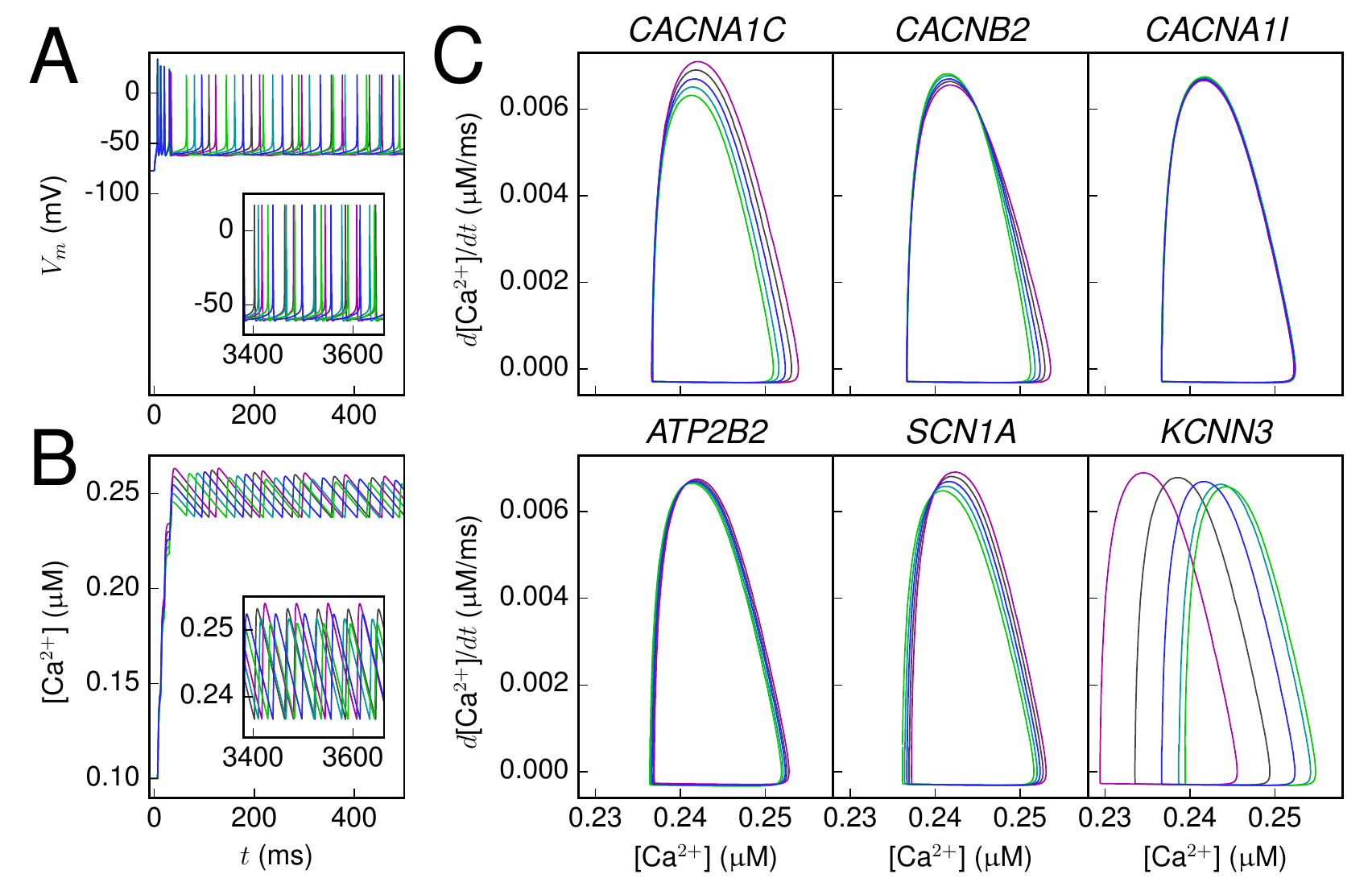}
\caption{
\textbf{Differences in \Cac limit cycles of the variants.}
\textbf{A,B}: The membrane potential (A) and \Ca concentration (B) time courses at soma as a response to a DC with an amplitude of 1.2 nA.
The different colors represent different scalings of a \CACNA1C variant \cite{kudrnac2009coupled} (scaling as in Figure \ref{fig1}).
Inset: The time course shown 3.4 s since the beginning of the stimulus.
\textbf{C}: The intracellular \Cac phase plane during steady firing caused by a DC applied to the soma. The variants and their scalings were chosen as in Figure \ref{fig1}C.
}
\label{fig3}
\end{figure}

We also studied the steady-state firing behavior by analyzing the limit cycle, i.e., the phase plane representation of the variable of interest after a large number of 
initial cycles. As the \Cac response served as the most sensitive biomarker, we used the \Cac limit cycle for this analysis. Figure \ref{fig3} shows, for some of the
gene variants, the \Cac limit cycle as a response to a long, square-pulse stimulus.

The different variants had different ways of impacting the shape of the \Cac limit cycle. An especially common trait was a shift or compression in the horizontal direction,
representing a small increase or decrease of intracellular \Cac during steady firing. Figure S1 shows the range of \Cac values observed during the constant
current injection for all variants used in this study. Interestingly, almost all variants gave deviations from the \Cac range observed in the control neuron.


The only genes that did not show notable variance for the range of \Cac during steady-state firing were \CACNA1I, \KCNS3, and \KCNB1. Despite this,
\CACNA1I did have a clear effect on the \Cac phase plane in the single-pulse response (see Figure \ref{fig1}C).
Conversely, the \KCNS3 and \KCNB1 variants were not found to influence any of the considered aspects of neuron behavior (see Figures S1 and S2).
In these variants, the activation and inactivation of the persistent \K current ($I_{Kp}$) have been modified. Our findings reflect the overall minor role that the current $I_{Kp}$ plays
in shaping the response properties of the L5PC model neuron: It is only present in the soma and with relatively small maximal conductance \cite{hay2011models}.
The minor contribution of this current to the model neuron behavior may be in contradiction with experimental evidence, which states that this current constitutes a major proportion
of the total outward \K currents \cite{guan2007kv2}.

\subsection*{Variants have an effect on integration of somatic and apical inputs}

A question that arises from the observation of the above trends is whether, and to what extent, the small deviations from control neuron behavior affect
the information processing capabilities of the neuron. 
In \cite{hay2011models}, it was shown that the L5PC model can be used to describe the \Ca spike generation
as a response to a combination of stimuli at the apical dendrite and at the soma. Moreover, they showed a good qualitative match with experimental data
and model predictions for the neuron responses during an ``up'' or ``down'' state.
In this work, we followed their definition for the down state as the resting state of the neuron, and the up state as a state where a subthreshold current
is applied to the proximal apical dendrite \cite{hay2011models}. We also adopted their protocol of combining an apical and somatic stimulus and studying the
effect of order and inter-stimulus interval (ISI) between the two (see Figure 8 in \cite{hay2011models}).
This gave us a temporal profile showing a range of ISI for which a large \Ca spike is produced. We then estimated the effect of our genetic variants on this
temporal profile.

\begin{figure}[!ht]
\centering
\includegraphics[clip,trim=50pt 473pt 55pt 45pt,width=0.6\textwidth]{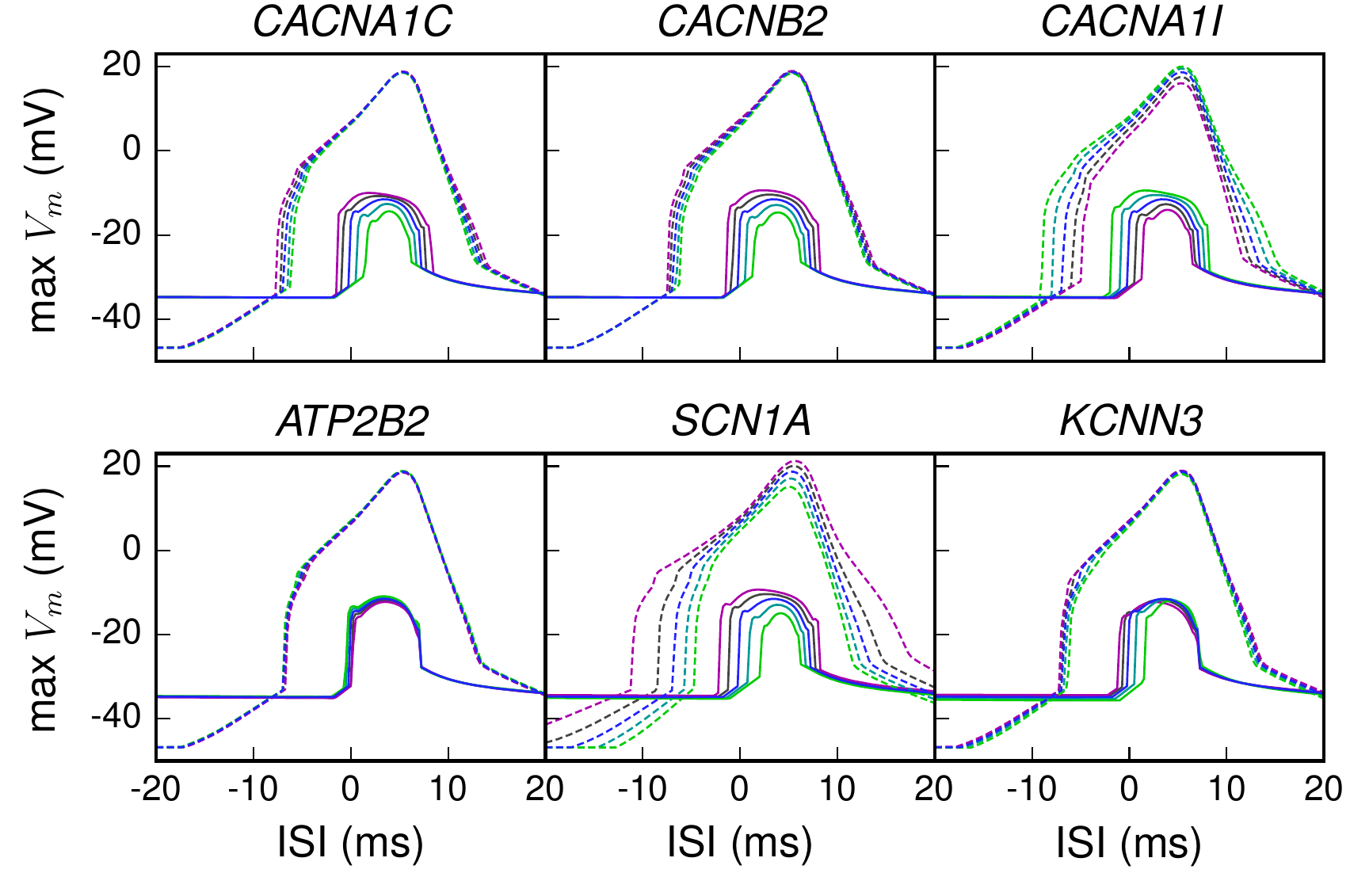}
\caption{
\textbf{Sensitivity of variant neurons to ISI between somatic and distal apical stimuli during up and down states.} 
The magnitude of the \Ca spikes was assessed by measuring the simulated membrane potential near the bifurcation point of the apical dendrite, i.e, at
a distance of 620 $\upmu$m from the soma. The dashed lines show the temporal maximum of the
model response membrane potential during a down state, and the solid lines during an up state. The x-axis represents the ISI between the somatic
and apical stimuli, positive values denoting cases where the apical stimulus was applied after the somatic stimulus. The variants shown are the same as in Figure \ref{fig1}. In
the up state paradigm, the neuron was first given a depolarizing square-pulse current of 0.42 nA $\times$ 200 ms at the proximal apical dendrite, 200 $\upmu$m from the soma.
In the middle of this period, a somatic square-pulse current of 0.5 nA $\times$ 5 ms was applied, and after a time defined as the ISI, an alpha-shaped current injection with
rise time 0.5 ms, decay time 5 ms, and maximal amplitude of 0.5 nA was applied. The down state paradigm was otherwise equal, but
the long depolarizing current was absent, and to compensate this, the short somatic square-pulse current had an amplitude of 1.8 nA.
}
\label{fig4}
\end{figure}

Figure \ref{fig4} shows the temporal profiles for the six representative variants. Effects of the different variants on the neuron response were visible
both during the down and during the up state. Particularly large effects were observed in \CACNA1I and \SCN1A variants, which code for the LVA \Ca channels and transient
\Na channels that contribute to the sharp rise of voltage during an action potential. These channel types are expressed both in the soma and in the apical dendrite.
Although all downscaled variants shared the basic form of the response curves, the shape was clearly altered in some variants.
As shown in \cite{hay2011models}, the up-state temporal profile of the control neuron showed elevated apical responses when the apical stimulus was applied
shortly after the somatic stimulus, but not when the apical stimulus was applied first. This order-specific response was changed in some of the variants
(see e.g. \SCN1A variant) so that they also produced large \Ca spikes when the apical stimulus was applied shortly before the somatic stimulus, and could hence alter
the order sensitivity of the coincidence detection.

\subsection*{Variants show differences in inhibition of a second apical stimulus}
The alteration of information processing by the variants can also be seen in the sensitivity to successive synaptic inputs at the same locations.
To explore this, we gave two successive stimuli to the apical dendrite, and varied the interval (ISI) between them.
The first stimulus activated long time scale
inhibitory currents, especially $I_{SK}$ and $I_m$, and within some following time window,
these currents could inhibit the generation of a subsequent spike.
This could be interpreted as a form of single-cell pre-pulse inhibition:
If the initial input caused the neuron to spike, the next input could fail to do so,
even at times when the second stimulus were of larger amplitude.

\begin{figure}[!ht]
\centering
\includegraphics[clip,trim=60pt 470pt 55pt 45pt,width=0.6\textwidth]{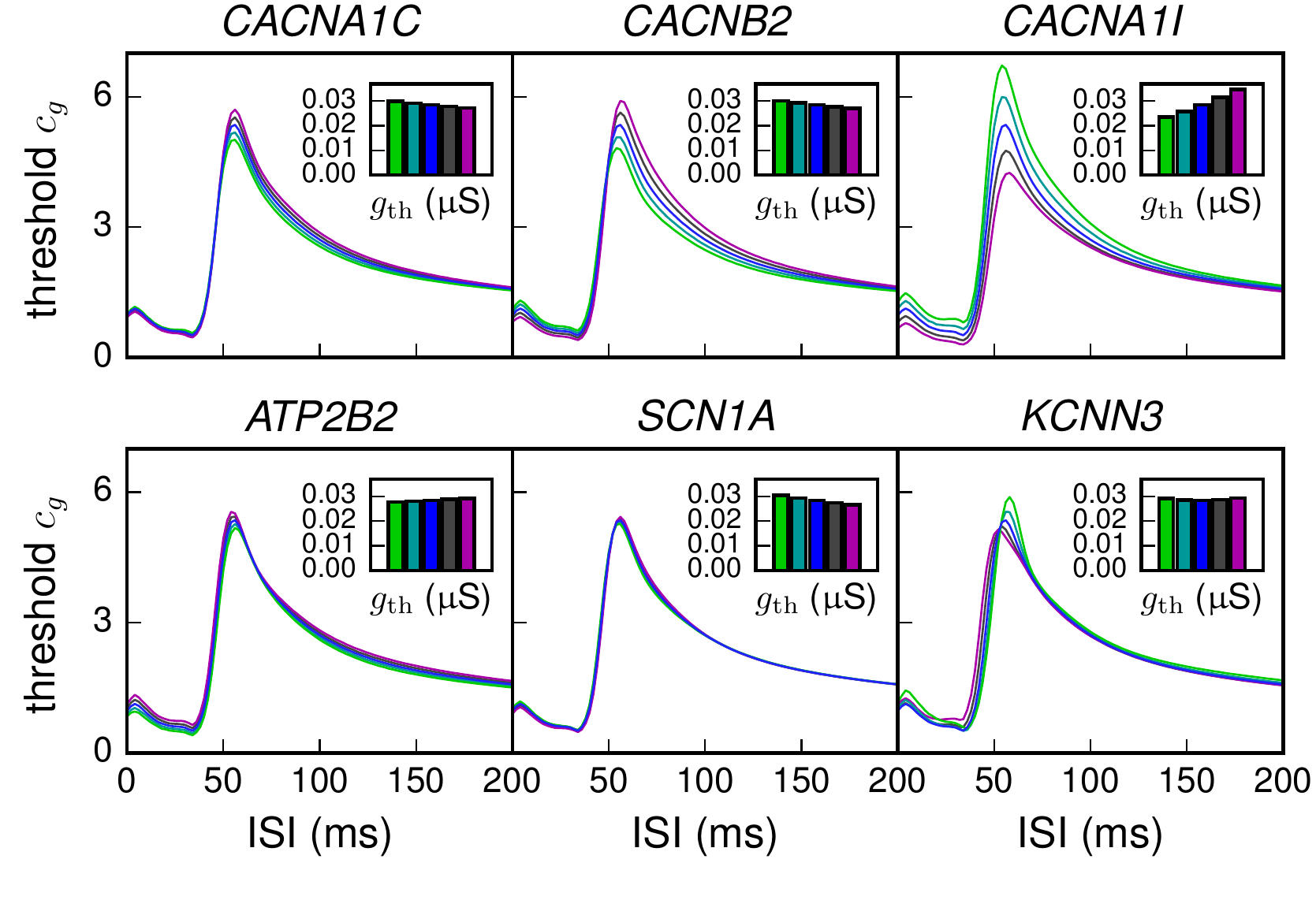}
\caption{
\textbf{Threshold conductance factor for inducing a second spike as a function of ISI.} The neuron was first made to spike by applying 3000 simultaneously activated
synaptic conductances that were uniformly distributed along the apical dendrite at a distance [300 $\upmu$m, $1300$ $\upmu$m] from the soma. All synaptic conductances were
alpha-shaped with $\tau=5$ ms and maximal conductance $1.1g_{\rm{th}}$, where $g_{\rm{th}}$ was the threshold conductance value for inducing a spike in resting state. These
values were variant-specific (and scaling-specific), and they are plotted in the inset of each panel. After the specified ISI, the synapses were activated again. The $y$-axis
shows the factor $c_g$, chosen such that $c_gg_{\rm{th}}$ was the threshold conductance of the second stimulus for generating an extra spike. Curves above the control curve
(blue) represent strengthened prepulse inhibition, while curves below it represent weakened prepulse inhibition. The effect of the \KCNN3 variant ($\epsilon=\frac 12$,
magenta) was bilateral: For ISI $\lesssim$ 50 ms, the effect was strengthening, whereas for ISI $\gtrsim$ 50 ms, the effect was weakening.
}
\label{fig5}
\end{figure}

Figure \ref{fig5} shows how the threshold synaptic conductance needed for inducing a second spike, relative to the one needed for the first spike, depends on
the ISI. Interestingly, the threshold conductance curve was affected by many of the variants, especially by the variants of \Ca channel genes.
Although the shown effect was partly attributed to the differences in the spiking thresholds between the variants, differences between the variants
could still be observed when an absolute spiking threshold values for inducing a second spike were considered (data not shown).
These findings raise the possibility that the disturbed prepulse inhibition observed
in many SCZ patients might be an effect of altered ionic channel properties at the single neuron level,
and not only of the modified elements in synaptic circuitry as previously thought \cite{swerdlow1993prepulse,medan2011dopaminergic} (cf. \cite{koch2012clinical}).

\subsection*{Combinations of variants may cause large alterations of neuron excitability}
The effects of the variants on neuron excitability, as explored in Figures \ref{fig1}--\ref{fig5}, were relatively small when the variants were studied in isolation 
(this was also the intention behind the downscaling process). However, combinations of variants of different genes may have additive effects on the neuron excitability.
As a final step of our inquiry, we went on to investigate the effect that combinations of several of the variants could have.
Figure S3 shows a selected example of how a combination of variants altered the neuron firing properties considered in Figures \ref{fig1}--\ref{fig5}.
Although this version of the model neuron consists of downscaled variants, the properties of neuron excitability were remarkably modified.

\section*{Discussion}

We presented a framework for studying the effects of genetic variants in genes related to schizophrenia (SCZ) on layer V pyramidal cell (L5PC) excitability.
Our results indicate that most of the studied variants predict an observable change in the neuron behavior
(see Figures \ref{fig1}--\ref{fig5}). Taken together, these data provide support for
using neuronal modeling to study the functional implications of SCZ-related genes on neuronal excitability. Although the analyses presented in this paper are specific to L5PCs
and SCZ-related genes, our framework may be directly applicable to other cell types and other polygenic diseases, such as bipolar disorder and autism,
given an identification of risk genes related to neuronal excitability. Furthermore, our approach can be directly extended to biophysically detailed models of neuronal networks, 
e.g. \cite{hay2014dendritic}.

The results shown in the present paper are based on a multi-compartmental neuron model \cite{hay2011models} employing a certain set of ionic channels. Alternative models with
comparable complexity have recently been published. The authors of \cite{bahl2012automated} present and apply a method for fitting the L5PC model to a simplified neuron morphology.
In another recent work \cite{almog2014quantitative}, the L5PC model was fit to each stained morphology and the corresponding recordings separately.
To rule out our conclusions being an artefact of a certain specific morphology, we carried out our analyses on
another neuron model fit (cell \#1 equipped with channel conductance values from the first column of Table S1 in \cite{hay2011models})
and an alternative neuron morphology (cell \#2 of \cite{hay2011models}). The trend of the variant effects remained the same in these simulations
(see Supplemental information, Figures S4 and S5). More comprehensive analysis could be carried
out by using the above-mentioned alternative models, or by employing a wider set of neuron morphologies \cite{hay2013preserving},
but this is left for future work.

Discovering the disease mechanisms of disorders with polygenic architecture, such as SCZ, requires considering the interplay between many different biological processes and entities.
The computational framework presented here allows one to identify physiological mechanisms and biomarkers common across multiple risk-related genes. Although there are many more SCZ-related
genes \cite{ripke2014biological} than those listed in Table \ref{tab_mutations}, the identified variants capture
multiple types of changes in neuron excitability (see Figures \ref{fig1}--\ref{fig5}). This allows studying the effects of combinations of different genetic
variants in a straightforward manner, as suggested in Supplemental information (Figure S3), although one should be careful in the interpretation of the results as certain variants
could have non-additive effects. The gene hits not considered in the present work contribute to e.g. expression level and function of ionotropic and G-protein coupled
receptors, as well as phosphorylation and dephosphorylation of diverse proteins. It should be noted that several of the identified SCZ-related genes, such as \textit{PTN},
\textit{PAK6}, and \textit{SNAP91} may also affect neuron morphology \cite{deuel2002pleiotrophin,nekrasova2008targeted,ang2003dock,dickman2006altered}. 
Our framework could be extended to include some of these effects, but studying the contribution of many signaling genes is out of the scope of the used neuron model.

Linking cellular mechanisms to SCZ can be extremely difficult due to the high cognitive level on which the disease symptoms usually emerge. There is however one important
exception, namely, altered prepulse inhibition, which is found in approximately 40-50\% of SCZ patients \cite{turetsky2007neurophysiological}.
In the present study, we showed that the considered genetic variants may have different tendencies to suppress a second synaptic stimulus, owing
only to the differences in the (non-synaptic) ion channel gating or \Ca dynamics (Figure \ref{fig5}). This could be an important contributor to understanding the phenomenon of prepulse
inhibition, although deficiencies in synaptic connections are most likely to play a role as well \cite{swerdlow1993prepulse}.
Moreover, altered integration of local and distal inputs in the manner of Figure
\ref{fig4} could possibly underlie incomplete or excessive activation of pyramidal cells in certain brain areas linked to auditory hallucinations (\cite{larkum2013cellular}; see also discussion of the role of
spontaneous activity in primary auditory cortex in triggering verbal hallucinations in \cite{kompus2013role}).
Experimental testing of the shown model predictions for integration of inputs on a single-cell level is possible as well. Even without genetic manipulations, the downscaled variants
of the ion channels could be imitated \textit{in vitro} by partial pharmacological blocks that are configured to cause changes in currents comparable to those in our
variants. However, in such a case the fine details on e.g. current activation time constants would be dismissed.

The genetic variation of ion channels will affect not only the membrane potentials of the neurons, but also the transmembrane currents generating the local field potential (LFP)
recorded inside cortex and the electrocorticography (ECoG) signals recorded at the cortical surface \cite{buzsaki2012origin, einevoll2013modelling}.
Further, the same transmembrane currents also determine the neuronal current dipole moments measured in EEG and MEG \cite{nunez2006electric,hamalainen1993magnetoencephalography}.
Thus a natural extension of the present work is to investigate the effects of the genetic variation of the single-neuron current
dipole moments \cite{linden2010intrinsic} and the corresponding single-neuron contribution to the EEG signals. This is left for future work.

A severe limitation of our approach is the lack of data on how different specific SCZ-related genetic polymorphisms affect the ion channel function and \Ca dynamics.
In this paper, we used literature data on extreme variants (e.g., loss of function mutations) of the identified SCZ-related risk genes, from a range of cell types,
including non-neural tissues. Ideally, one would want to use physiological measures for the relevant genetic variants in the actual cell types (i.e., L5PCs)
in cortical circuits, but such data do not currently exist. On the other hand, the diversity of data in the literature that we used and the possible discrepancies in the experimental procedures 
therein could also reflect in the results shown in this study, although our downscaling procedure constrains such errors from above. Novel methods employing automated cell patching
\cite{ranjan2014automated} to test the effects of subtle SCZ-related SNPs could provide the solution to both above limitations of our study. 

The validity of the downscaling, which for the aforementioned reasons is a central procedure in our framework, is built upon two assumptions: 1) That the SCZ-related gene variants do affect
neuronal excitability on a single-neuron level, and 2) that their effects on ion channel or calcium pump kinetics are smaller than those of certain already documented mutations with extreme
phenotypic consequences, such as deafness, hemiplegic migraine, or other pain disorders. The first assumption can be argued for by the multitude of ion channel-encoding genes that have been shown to be
related to SCZ risk, and the second by the lack of severe body malfunctioning (such as the abovementioned ones) in SCZ patients, but rigorous proofs for both assumptions are yet to be shown.
A challenge for future studies is also to decipher how the ion channel densities are affected by the variants in various cell types, an issue that might be important
in modeling the effects of SCZ-associated polymorphisms. 

Our framework is an early attempt toward understanding the disease mechanisms of polygenic psychiatric disorders by computational means.
It binds together genetic and single neuron levels of abstraction in the big picture of modeling psychiatric illnesses, as recently sketched in
\cite{wang2014computational} and \cite{corlett2014computational}.
While earlier computational studies considered only effects of single variants on neuron firing behavior \cite{murbartian2004functional,zhang2011expression},
our framework allows the screening of multiple types of variants and their implications on the neuron excitability.
It may serve as a proof of principle for a novel ``Biophysical Psychiatry'' approach, enabling the integration of information from previously disparate
fields of study including GWASs, functional genomics, and biophysical computational modeling using models of moderate complexity.
We consider the chosen level of complexity a suitable trade-off for this purpose: While more approximate models, such as integrate-and-fire models,
could not distinguish between the effects of different genetic variants in an acceptable precision, more detailed models including e.g. dynamics of
protein translation and phosphorylation are likely to be very computation-intensive.
The challenge for future work is to extend the study to both larger and finer spatial and temporal scales.

\section*{Acknowledgements}
NOTUR resources (NN4661K, NN9114K, NS9114K) were used for heavy simulations. Funding: NIH grant 5 R01 EB000790-10, EC-FP7 grant 604102 (``Human Brain Project''),
Research Council of Norway (216699, 213837 and 223273), South East Norway Health Authority (2013-123), and KG Jebsen Foundation. 

\section*{Financial disclosures}
The authors declare that there are no conflicts of interest.

\section*{Author contributions}
Designed the study: TMM, GTE, OAA, AMD.
Provided data or analytical support: SD, GH, AD, YW, GTE, AMD.
Performed the analysis: TMM.
Interpreted the results: TMM, GH, AD, AW, FB, SD, YW, GTE, OAA, AMD.
Wrote the manuscript: TMM, GH, AD, GTE, OAA, AMD.

\bibliographystyle{apalike}
\bibliography{scz_abbr}


\includepdf[pages={1-18}]{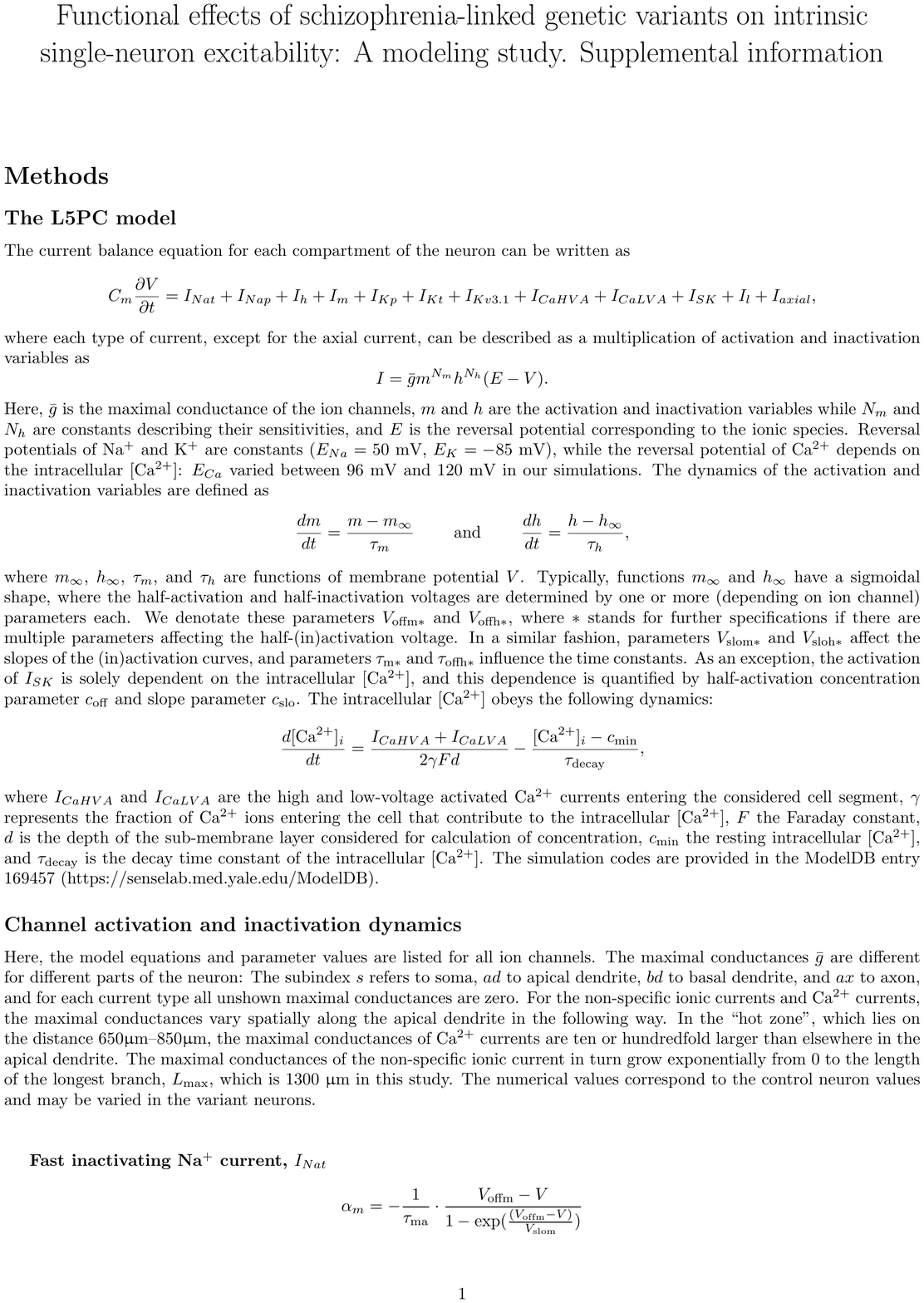}

\end{document}